# Why "AI" Models for Predicting Soil Liquefaction have been Ignored, Plus Some that Shouldn't Be

Brett W. Maurer, M.EERI[1] and Morgan D. Sanger, M.EERI[1]

**Abstract**

Soil liquefaction remains an important and interesting problem that has attracted the development of enumerable prediction models. Increasingly, these models are utilizing algorithmic learning, or "artificial intelligence" (AI). The rapid growth of AI in the liquefaction literature is unsurprising, given its ease of implementation and potential advantages over traditional statistical methods. However, AI liquefaction models have been widely ignored by practitioners and researchers alike; the objective of this paper is to investigate "why?" Through a sample review of 75 publications, we identify several good reasons. Namely, these models frequently: (i) are not compared to state-of-practice models, making it unclear why they should be adopted; (ii) depart from best practices in model development; (iii) use AI in ways that may not be useful; (iv) are presented in ways that overstate their complexity and make them unapproachable; and (v) are discussed but not actually provided, meaning that no one can use the models even if they wanted to. These prevailing problems must be understood, identified, and remedied, but this does not mean that AI itself is problematic, or that all prior efforts have been without merit or utility. Instead, understanding these recurrent shortcomings can help improve the direction and perceptions of this growing body of work. Towards this end, we highlight papers that are generally free from these shortcomings, and which demonstrate applications where AI is more likely to provide value in the near term: permitting new modeling approaches and potentially improving predictions of liquefaction phenomena.

**Introduction**

As one of the most complex, controversial, and consequential phenomena in geotechnical engineering, soil liquefaction has garnered significant research investment over the last half-century. This includes the development of enumerable models for predicting the occurrence and/or consequences of liquefaction. Collectively, these models use inputs that vary in quantity and complexity and are based on approaches that range from largely empirical to largely mechanistic. More recently, and mirroring the scientific and business communities at large, these models often employ algorithmic learning, which is more commonly known as "narrow artificial intelligence," or "AI." The fervent adoption of AI into the liquefaction literature is unsurprising, given both the ease of developing AI models and their potential advantages over "traditional" statistical methods, especially when given large datasets. We now estimate, for example, that most models published to date for predicting liquefaction triggering have utilized AI. It is thus striking – perhaps – that these models have been almost universally ignored, both by the practicing geotechnical community and by many liquefaction researchers. Consider, for example, that one state-of-practice (henceforth "SOP") liquefaction triggering model – that of Youd et al. (2001) – has 3,183 citations at the time of our writing, whereas 75 other liquefaction models based on AI and subsequently reviewed herein have a *total* of 2,425 citations. Similarly striking comparisons can be made with other former and current SOP models (e.g., Seed and Idriss, 1971; Robertson and Wride, 1998; Andrus and Stokoe, 2000; Cetin et al., 2004; Boulanger and Idriss, 2014; etc.). This disparity is not simply a matter of model maturity, as illustrated by Figure 1. Here, the citation-per-year counts are shown for 10 SOP liquefaction response models (i.e., popular models that predict liquefaction without AI); these are juxtaposed with the counts for 75 AI models. On average, the 10 SOP models each garner 76 citations per year, whereas the 75 AI models average fewer than 4. Of course, these models are in in some sense not independent; SOP models have been incrementally built from a common framework that is widely used, trusted, and endorsed by expert consensus (e.g., Youd et al., 2001), whereas AI models are more dissimilar and have yet to receive any such endorsement, let alone be widely discussed. Anecdotally, we are unaware of a single use-case of an AI liquefaction model in geotechnical practice, and are similarly unaware of any building code, design manual, or textbook that explicitly permits, encourages, or even mentions the use of AI liquefaction models. In effect, these models have been used by almost no one other than their respective developers. The overarching objective of this paper is to investigate the question of "why?"

---

[1]Department of Civil & Environmental Engineering University of Washington, Seattle, WA

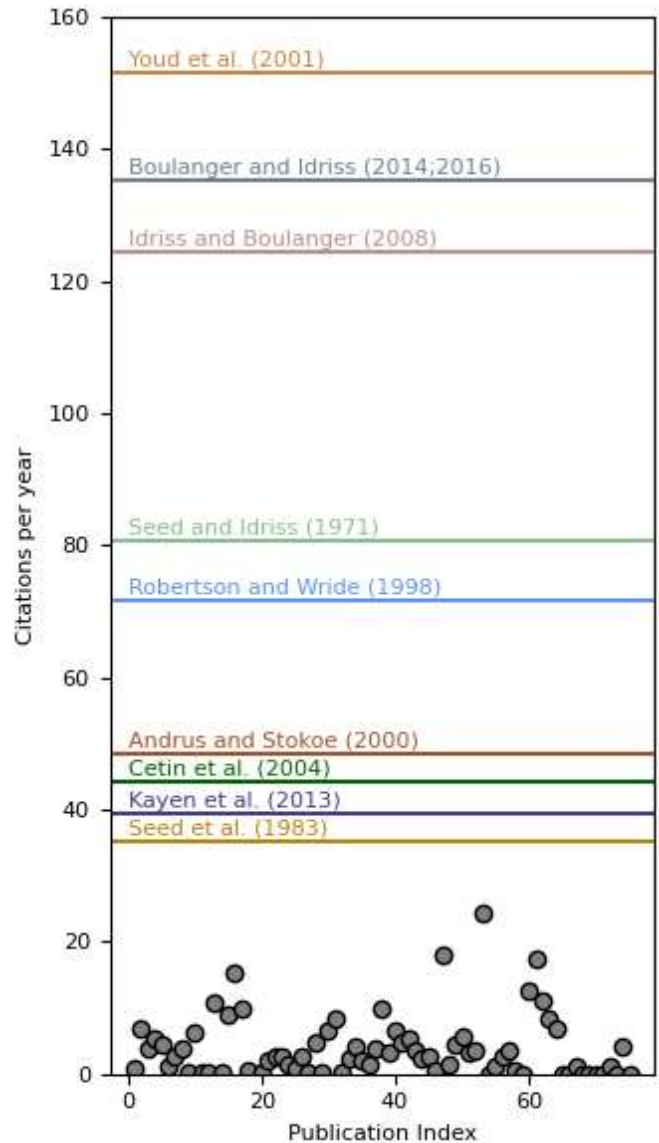

**Figure 1.** Citations/year for 10 SOP and 75 AI liquefaction models reviewed herein; Publication Index = AI papers 1 through 75 ordered chronologically, as listed in Table 1 (statistics from Google Scholar).

We will argue, via a review of 75 published papers, that there are good reasons for the general dismissal of existing AI liquefaction models. Namely, these models frequently: (i) are not compared to SOP models, making it unclear why they should be adopted; (ii) depart from various best practices with respect to data, model training, and performance evaluation; (iii) use AI in ways that are unlikely to be useful; (iv) are presented in ways that overstate their complexity and make them unapproachable, veiling the lack of true innovation (i.e., the "woo-woo" effect); and (v) are discussed but not actually provided, meaning that no one can use the models even if they wanted to. Although these practices are recurrent, problematic, and must be ceased, they are generally not shortcomings of AI itself. Yet with numerous AI liquefaction models flooding the literature each year, it can be difficult to discern those that are potentially innovative from those that are not. An unfortunate outcome emerges from this landscape: the widespread dismissal of AI by some in the geotechnical community. Anecdotally, we have heard from researchers and practitioners for whom "AI" has negative connotations, and who tend to group and dismiss AI models wholesale. While we understand this mindset, we deem it a mistake. Empirical and semi-empirical models and model components are ubiquitous in liquefaction modeling. If such models are effective – as their widespread use suggests – it follows that AI can play a role in the field, just as it increasingly does in nearly all fields of empirical inquiry.

Accordingly, this paper discusses the prevailing problems with AI liquefaction models and papers so that they may be understood and identified, and so that works meriting greater consideration may be distinguished from those that superficially seem similar. Towards that end, we highlight several of the reviewed papers that are free from the five shortcomings summarized previously and subsequently discussed. We also contrast areas of liquefaction research where AI has been useful with areas where AI has arguably not been deployed as sensibly. This work partly stems from that of Xie et al. (2020), who reviewed nearly 200 AI publications across all of earthquake engineering, including 25 pertinent to liquefaction. Notably, Xie et al. (2020) concluded that AI liquefaction models "outperform" SOP models, including Boulanger and Idriss (2014). We consider this conclusion misleading, but also reflective of the confusion that surrounds the topic. Aside from Xie et al. (2020), we are unaware of any commentary or comprehensive review of AI in liquefaction modeling, which stands in contrast to the large and growing literature on the subject.

In the following, a short overview of the 75 papers is given and the papers are contextualized in the domains of liquefaction research and AI tools. The common categorical shortcomings of AI papers are then presented and backed by statistics from the literature. Lastly, areas of research where AI is more likely to make significant contributions are discussed, and representative papers are highlighted.

**Literature Overview and Background**

A total of 75 publications were reviewed, including 65 journal papers and 10 conference papers published from 1993 to the present. This review was intended to sample the breadth of the existing literature (i.e., to not miss any unique uses of AI or novel methodologies) but is not necessarily exhaustive. The interested reader will find additional publications on the subject, and more will have appeared since our review. The distributions of reviewed publications by outlet and by year are shown in Figures 2a and 2b, respectively, and a complete listing of the titles is given in Table 1. Of the 75 publications, the most appeared in *Soil Dynamics and Earthquake Engineering* (14), followed by *Computers and Geotechnics* (5) and *Natural Hazards* (5). Papers from other relatively high-impact factor journals, including the *Journal of Geotechnical and Geoenvironmental Engineering*, *Engineering Geology*, *Earthquake Spectra*, *Soils and Foundations*, the *Canadian Geotechnical Journal*, and *Bulletin of Earthquake Engineering*, were also reviewed. Ten papers appeared at conferences, including several in ASCE proceedings. Many well-known outlets of geotechnical engineering research are thus represented. As seen in Figure 2b, the number of papers on the subject appears to be increasing annually, with 5 in the 1990s, 16 in the 2000s, and 26 in the 2010s; the 2020s are currently on pace for more than 80 such papers. It is thus timely to address this body of work, and ideally, to shape its direction and perceptions for the better.

Maurer, B. W., & Sanger, M. D. (2023). Why "AI" models for predicting soil liquefaction have been ignored, plus some that shouldn't be. *Earthquake Spectra*, 87552930231173711.

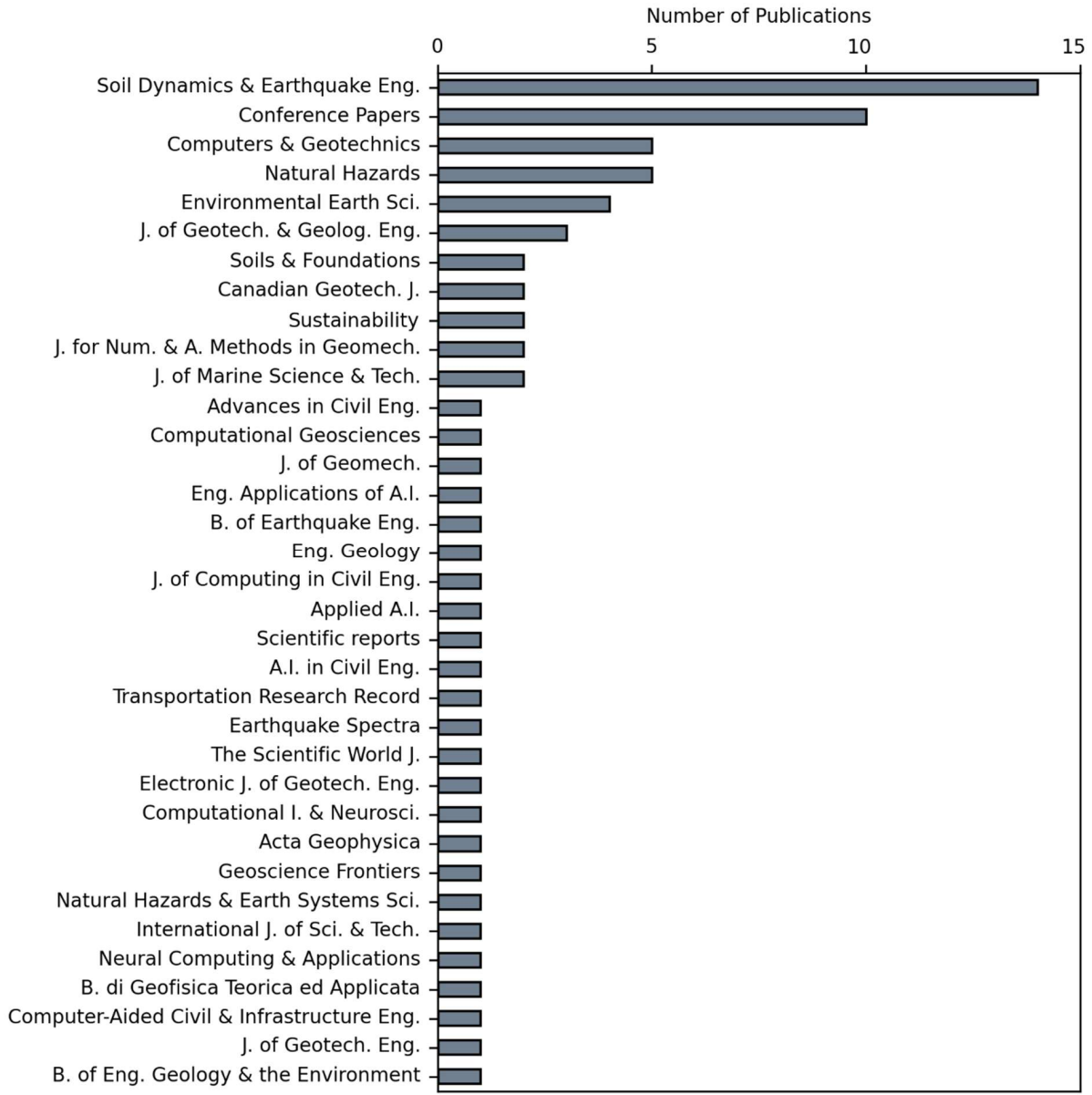

(a)

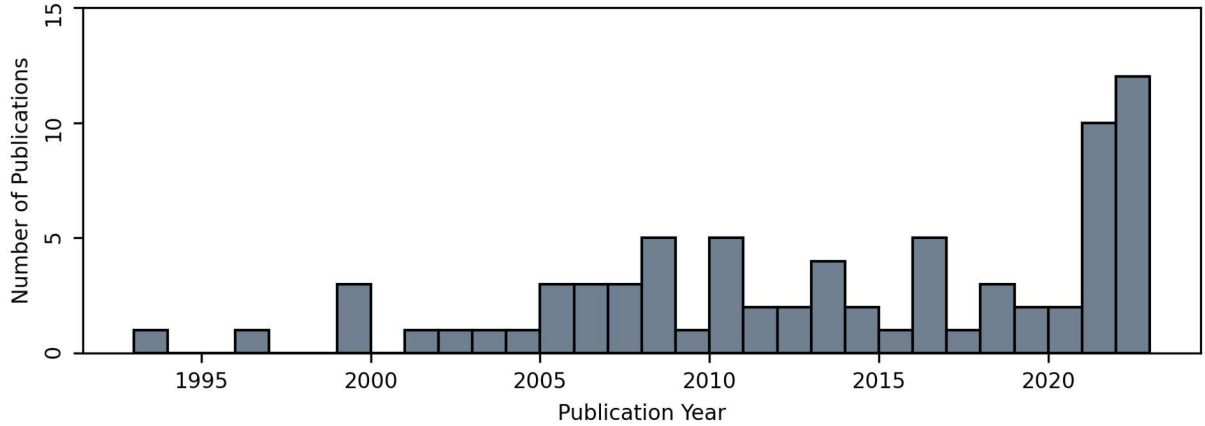

**(b)**

**Figure 2.** Quantities of publications reviewed per: **(a)** outlet; and **(b)** publication year.

Table 1. Literature summary (some titles abbreviated; see references for full citations).

| Publication Index | Reference | Title of Publication |
|---|---|---|
| 1 | Tung et al. (1993) | Assessment of liquefaction potential using neural networks |
| 2 | Goh (1996) | Neural-network modeling of CPT seismic liquefaction data |
| 3 | Hsein Juang et al. (1999) | Appraising cone penetration test-based liquefaction resistance evaluation methods: artificial neural network approach |
| 4 | Juang & Chen (1999) | Cpt-based liquefaction evaluation using artificial neural networks |
| 5 | Wang & Rahman (1999) | A neural network model for liquefaction-induced horizontal ground displacement |
| 6 | Chiru-Danzer et al. (2001) | Estimation of liquefaction-induced horizontal displacements using artificial neural networks |
| 7 | Rahman & Wang (2002) | Fuzzy neural network models for liquefaction prediction |
| 8 | Baziar & Nilipour (2003) | Evaluation of liquefaction potential using neural-networks and CPT results |
| 9 | Hao et al. (2004) | Evaluation of sands liquefaction potential based on SOFM neural network |
| 10 | Baziar & Ghorbani (2005) | Evaluation of lateral spreading using artificial neural networks. |
| 11 | Garg (2005) | Evaluation of liquefaction potential using adaptive resonance theory based neural networks |
| 12 | Kurup & Garg (2005) | Evaluation of liquefaction potential using neural networks based on adaptive resonance theory |
| 13 | Javadi et al. (2006) | Evaluation of liquefaction induced lateral displacements using genetic programming |
| 14 | Liu et al. (2006) | Artificial neural network methodology for soil liquefaction evaluation using CPT values |
| 15 | Pal (2006) | Support vector machines-based modelling of seismic liquefaction potential |
| 16 | Goh & Goh (2007) | Support vector machines: their use in geotechnical engineering as illustrated using seismic liquefaction data |
| 17 | Hanna et al. (2007) | Neural network model for liquefaction potential in soil deposits using Turkey and Taiwan earthquake data |
| 18 | Khozaghi & Choobbasti (2007) | Predicting of liquefaction potential in soils using artificial neural networks |

Maurer, B. W., & Sanger, M. D. (2023). Why "AI" models for predicting soil liquefaction have been ignored, plus some that shouldn't be. *Earthquake Spectra*, 87552930231173711.

| | | |
|---|---|---|
| 19 | Chen et al. (2008) | Empirical model for liquefaction resistance of soils based on artificial neural network learning of case histories |
| 20 | Chern et al. (2008) | CPT-based liquefaction assessment by using fuzzy-neural network |
| 21 | Garcia et al. (2008) | A neurofuzzy system to analyze liquefaction-induced lateral spread |
| 22 | Livingston et al. (2008) | Using Decision-Tree Learning to Assess Liquefaction Potential from CPT and Vs |
| 23 | Ramakrishnan et al.(2008) | Artificial neural network and liquefaction susceptibility assessment: a case study using the 2001 Bhuj earthquake data, Gujarat, India |
| 24 | Chern & Lee (2009) | CPT-based simplified liquefaction assessment by using fuzzy-neural network |
| 25 | Hao & Chen (2010) | AFSs-RBF neural network for predicting earthquake-induced liquefaction of light loam |
| 26 | Oommen & Baise (2010) | Model development and validation for intelligent data collection for lateral spread displacements |
| 27 | Rahman et al. (2010) | Artificial neural network in CPT base liquefaction prediction |
| 28 | Rezania et al. (2010) | Evaluation of liquefaction potential based on CPT results using evolutionary polynomial regression |
| 29 | Samui & Sitharam (2010) | Relevance vector machine for evaluating seismic liquefaction potential using shear wave velocity |
| 30 | Rezania et al. (2011) | An evolutionary based approach for assessment of earthquake-induced soil liquefaction and lateral displacement |
| 31 | Samui & Sitharam (2011) | Machine learning modelling for predicting soil liquefaction susceptibility |
| 32 | García et al. (2012) | Liquefaction Assessment through Machine Learning Approach |
| 33 | Liu & Tesfamariam (2012) | Prediction of lateral spread displacement: data-driven approaches |
| 34 | Gandomi et al. (2013) | Decision tree approach for soil liquefaction assessment |
| 35 | Samui & Karthikeyan (2013) | Determination of liquefaction susceptibility of soil: a least square support vector machine approach |
| 36 | Venkatesh et al. (2013) | Appraisal of liquefaction potential using neural network and neuro fuzzy approach |
| 37 | Xue & Yang (2013) | Application of the adaptive neuro-fuzzy inference system for prediction of soil liquefaction |
| 38 | Goh & Zhang (2014) | An improvement to MLR model for predicting liquefaction-induced lateral spread using multivariate adaptive regression splines |
| 39 | Muduli & Das (2014) | Evaluation of liquefaction potential of soil based on standard penetration test using multi-gene genetic programming model |
| 40 | Kohestani et al. (2015) | Evaluation of liquefaction potential based on CPT data using random forest |
| 41 | Abbaszadeh Shahri (2016) | Assessment and prediction of liquefaction potential using different artificial neural network models: a case study |
| 42 | Hu et al. (2016) | Assessment of seismic liquefaction potential based on Bayesian network constructed from domain knowledge and history data |
| 43 | Kaya (2016) | Predicting liquefaction-induced lateral spreading by using neural network and neuro-fuzzy techniques |
| 44 | Samui et al. (2016) | An alternative method for determination of liquefaction susceptibility of soil |
| 45 | Xue & Xiao (2016) | Application of genetic algorithm-based support vector machines for prediction of soil liquefaction |
| 46 | Kumar & Rawat (2017) | Soil liquefaction and its evaluation based on SPT by soft-computing techniques |
| 47 | Hoang & Bui (2018) | Predicting earthquake-induced soil liquefaction based on a hybridization of kernel Fisher discriminant analysis and a least squares support vector machine: a multi-dataset study |
| 48 | Nejad et al. (2018) | Evaluation of liquefaction potential using random forest method and shear wave velocity results |
| 49 | Pirhadi et al. (2018) | A new equation to evaluate liquefaction triggering using the response surface method and parametric sensitivity analysis |
| 50 | Alobaidi et al. (2019) | Predicting seismic-induced liquefaction through ensemble learning frameworks |
| 51 | Kurnaz & Kaya (2019) | A novel ensemble model based on GMDH-type neural network for the prediction of CPT-based soil liquefaction |
| 52 | Das et al. (2020) | Multi-objective feature selection (MOFS) algorithms for prediction of liquefaction susceptibility of soil based on in situ test methods |
| 53 | Njock et al. (2020) | Evaluation of soil liquefaction using AI technology incorporating a coupled ENN/t-SNE model |
| 54 | Alzahamie & Abdul-Husain (2021) | Artificial neural network for prediction of liquefaction triggering based on CPT data |
| 55 | Babacan & Ceylan (2021) | Evaluation of soil liquefaction potential with a holistic approach: a case study from Araklı |
| 56 | Durante & Rathje (2021) | An exploration of the use of machine learning to predict lateral spreading |
| 57 | Kumar et al. (2021) | A Novel Methodology to Classify Soil Liquefaction Using Deep Learning |
| 58 | Pham (2021) | Application of feedforward neural network and SPT results in the estimation of seismic soil liquefaction triggering |
| 59 | Wang & Wang (2021) | Sandy Soil Liquefaction Prediction Based on Clustering-Binary Tree Neural Network Algorithm Model |
| 60 | Zhang & Wang (2021) | An ensemble method to improve prediction of earthquake-induced soil liquefaction: a multi-dataset study |
| 61 | Zhang et al. (2021) | The adoption of ELM to the prediction of soil liquefaction based on CPT |
| 62 | Zhang et al. (2021) | The adoption of a support vector machine optimized by GWO to the prediction of soil liquefaction |
| 63 | Zhao et al. (2021) | A novel PSO-KELM based soil liquefaction potential evaluation system using CPT and Vs measurements |
| 64 | Demir & Sahin (2022) | Comparison of tree-based machine learning algorithms for predicting liquefaction potential using canonical correlation forest, rotation forest, and random forest based on CPT data |
| 65 | Demir & Şahin (2022) | Liquefaction prediction with robust machine learning algorithms (SVM, RF, and XGBoost) supported by genetic algorithm-based feature selection and parameter optimization from the perspective of data processing |
| 66 | Fahim et al. (2022) | Liquefaction resistance evaluation of soils using artificial neural network for Dhaka City, Bangladesh |
| 67 | Geyin et al. (2022) | An AI driven, mechanistically grounded geospatial liquefaction model for rapid response and scenario planning |
| 68 | Guo et al. (2022) | Soil liquefaction assessment by using hierarchical Gaussian Process model with integrated feature and instance based domain adaption for multiple data sources |
| 69 | Hanandeh et al. (2022) | A Comparative Study of Soil Liquefaction Assessment Using Machine Learning Models |
| 70 | Ozsagir et al. (2022) | Machine learning approaches for prediction of fine-grained soils liquefaction |
| 71 | Rateria & Maurer (2022) | Evaluation and updating of Ishihara's (1985) model for liquefaction surface expression, with insights from machine and deep learning |
| 72 | Todorovic & Silva (2022) | A liquefaction occurrence model for regional analysis |
| 73 | Zhang et al. (2022) | Soil Liquefaction Prediction Based on Bayesian Optimization and Support Vector Machines |
| 74 | Zhao et al. (2022) | CPT-based fully probabilistic seismic liquefaction potential assessment to reduce uncertainty: Integrating XGBoost algorithm with Bayesian theorem |
| 75 | Jena et al. (2023) | Earthquake-induced liquefaction hazard mapping at national-scale in Australia using deep learning techniques |

Each of the 75 papers developed at least one AI model for predicting the occurrence and/or a consequence of liquefaction (e.g., ejecta, lateral spreading). To place these models in context, it will be useful to briefly define the hierarchy of SOP models, which we view as having three tiers. The first tier, "tier 1," is defined by "geospatial" models (e.g., FEMA, 2013; Zhu et al., 2017), which predict ground failure using existing mapped information and above-ground inferences of below-ground conditions, generally at broad spatial resolutions. Relative to other models, those in tier 1 use data that is less site-specific, less invasive, and more freely available. The second tier, "tier 2," includes semi-empirical "stress-based" and "energy-based" models (e.g., among many, Green and Mitchell, 2004; Moss et al., 2006; Kayen et al. 2013; Kokusho, 2013; Boulanger and Idriss, 2014; Cetin et al., 2018; Green et al., 2019; Saye et al., 2021), which use in-situ geotechnical data and mechanistic principles to predict liquefaction response for one-dimensional profiles. Of these, "stress-based" models are more common and are synonymous with "triggering curves" that identify combinations of cyclic stress ratio (CSR) and cyclic resistance ratio (CRR) that are, and are not, expected to result in liquefaction, wherein CRR is most often predicted from standard penetration test (SPT) blow count, cone penetration test (CPT) tip resistance, or shear wave velocity ($V_s$). This second tier can also be considered to include models that predict manifestations of liquefaction, and which serve as extensions to triggering models. Because triggering refers to liquefaction at discrete depths in a profile, the outputs from triggering analysis are often used in series with manifestation models to predict a profile's system response in the form of settlement, ejecta, spreading, cracking, etc. (e.g., Iwasaki, 1978; Zhang et al., 2002; Youd et al., 2002; Maurer et al., 2015; Hutabarat and Bray, 2022). The third tier, "tier 3," is defined by "constitutive" models that predict liquefaction response at the soil element level, typically using numerous input parameters. Tier 3 models, which are intimately linked to "numerical" methods (e.g., the finite element method), aim to account for mechanics more completely and continuously in both time and space, yet are still empirical in many respects.

Of these, tier 2 models are by far the most popular in routine practice, and as such, it's not surprising that ~95% of the 75 AI models address this tier. That is, they aim to predict liquefaction triggering (~75% of models) and/or its consequences (~20% of models) using the same types of data as tier-2 SOP methods. The remaining ~5% of the AI models can be classified as tier 1, meaning that they predict liquefaction over regional or global extents. Although tier 3 AI constitutive models have begun to appear, they are relatively rare and were not considered in our analysis, given that these models are functionally very different from tier 1 and 2 models. Other models that predict laboratory cyclic test results (e.g., pore pressure generation) and which are most closely aligned with tier 3, were similarly omitted from the present scope.

It will also be useful to succinctly summarize AI methods, both to place the tools used by the 75 papers in context, and to frame our discussions of where AI is, and is not, likely to provide value. However, given the countless references on AI methods in the literature, we opt not to recapitulate their mathematical details and instead focus our summary on how AI workflows and models tend to differ from more traditional techniques. Algorithmic learning is sometimes parsed into "machine learning" and "deep learning." Machine learning techniques include but are not limited to: (i) decision trees (DTs), which are built and ensembled using processes such as bagging, boosting, and random forests; (ii) support vector machines (SVMs); (iii) gaussian process models (GPMs); (iv) k-nearest neighbors (KNN); and (v) generalized additive models (GAMs). Deep learning generally refers to artificial neural networks (ANNs), specifically those with multiple "layers."

Most AI techniques can be applied both to class problems (e.g., liquefaction triggering) and to continuous problems, or in AI lexicon – "regression" problems (e.g., lateral spreading displacement). All can be used to train models that map some combination of inputs to a target output. Each model has internal options (called "hyperparameters"), that affect a model's form and function (not to be confused with its predictor variables). Once promising models are identified, "hyperparameter optimization" is commonly used to train model variants in search of the most optimal. Following their development, multiple models are often ensembled to form a model that is more stable and generalizable (e.g., likely to perform better on unfamiliar data) compared to any one base predictor. Ensembling can be as simple as a traditional logic tree or as complex as "stacking" (or "meta-learning"), which is AI modeling using other AI models as inputs.

Shown in Figure 3 is the distribution of techniques used by the reviewed papers. Most (~77%) use some form of neural network, followed in popularity by decision trees and support vector machines. The preference for neural networks is notable because these models generally: (i) require larger datasets to train;

and (ii) are not interpretable, meaning the model's behavior cannot be well explained. It should also be noted that a small number of AI papers used "other" techniques that include linear and logistic regression, which respectively predict scores and classes. Because AI refers less to a specific set of tools and more to an algorithmic (say, automated) approach, it can include these "traditional" statistical methods, and indeed many AI libraries do. Therefore, it's worth discussing how AI methods tend to differ from the methods traditionally used to train liquefaction models, given that these methods have considerable overlap.

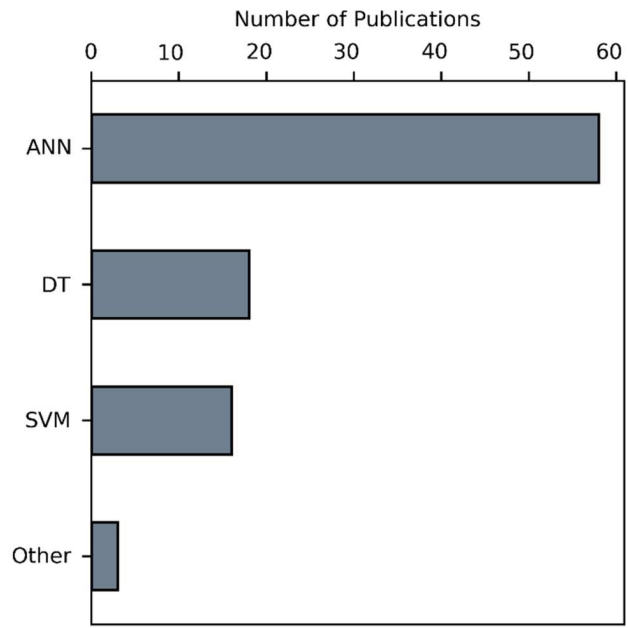

**Figure 3.** Distribution of AI techniques among reviewed publications; ANN = artificial neural network; DT = decision tree; SVM = support vector machine.

First, AI is based in the pragmatic. It is concerned more with predictive power and less with understanding a problem's governing fundamentals. Some AI models are so convoluted that their predictions are nearly impossible to explain. Others are interpretable (or "explainable" in AI vernacular), which generally means that a variable's contribution to a prediction can be determined (i.e., the relative importance of predictors may be ranked). The precise relationships between predictors and targets, however, almost always remain unknown. There are many problems in business, as in geotechnics, where the "governing fundamentals" are wholly abstract, yet the accurate predictions speak for themselves. No one expects, for example, that theoretical models should exist to select online ads, recognize images, or transcribe speech. "Alexa," "Siri," and other such models understand speech without understanding language. They learn from examples, and arguably, there are many analogs in geotechnical engineering. Would a mechanistic model relating SPT and CPT measurements be helpful? Probably, but most engineers would be content to perfectly predict one from the other regardless of how the model is made. Will a theoretical relationship between topographic slope and subsurface $V_s$ ever be discovered? Definitely not (the very question is abstract), but one is quite useful for predicting the other.

Second, and relating to the first, AI models tend *not* to be portable and executable by hard copy. AI models require code, either in raw form or embedded in a user interface (e.g., a Windows executable program). Some model predictions could possibly be stored (say, in a geographic map or lookup table), but AI models themselves can very rarely be presented in an equation. If code or software is not provided, no one can use the model. Third, AI modelers tend to work with more explanatory variables – in some cases hundreds – and variables of different types, as compared to traditional statisticians. Traditional regression fundamentally requires hypotheses of what matters and how (i.e., which liquefaction predictors are useful and how they relate to the outcome). This means that far more emphasis must be placed on manually selecting variables, which increases the likelihood that useful predictors and/or complex interactions amongst variables will be overlooked. AI allows for a very large number of prospective variables to be used, with greater potential for the predictive information stored in those variables to be exploited.

Fourth, and extending from the third, the use of many predictor variables creates additional considerations. First, it tends to necessitate more data. Although AI models can outperform others, this is predicated on there being enough samples to learn from (as in thousands, if not millions of data points). This is particularly true of neural networks. If a large dataset is unavailable, simpler models may perform as well or better. Additionally, the use of many predictor variables tends to increase the possibility of overfitting (i.e., where a model performs much better on training data than unbiased test data). Although overfitting is a risk for any model, methods to assess and mitigate it, such as training-testing data splitting and k-fold cross-validation, should be routine in AI workflows but are less common in traditional regression.

As a summary example, consider the classic Youd et al. (2002) model for lateral spreading displacement on sloping ground, as defined by Eq. (1):

$$\log D_H = -16.213 + 1.532M - 1.406 \log R^* - 0.012R + 0.338 \log S + 0.540 \log T_{15} + 3.413 \log(100 - F_{15}) - 0.795 \log(D50_{15} + 0.1mm) \tag{1}$$

Where: $D_H$ is the predicted lateral ground displacement (m); $M$ is moment magnitude; $R^* = R + 10^{(0.89M-5.64)}$, where $R$ is the map distance to the seismic energy source (km); $S$ is the ground slope (%); $T_{15}$ is the cumulative thickness (m) of saturated, granular layers with corrected blow count ($N_{160}$) less than 15; $F_{15}$ is the average fines content (fraction) for granular materials in $T_{15}$; and $D50_{15}$ is the average mean grain size (mm) of granular material in the $T_{15}$.

The Youd et al. (2002) model is not far removed from AI in some respects. Yet in general, an AI modeler would likely consider many more predictor variables than Youd et al. (2002) did and could do so far more easily. That modeler would develop multiple types of models and numerous variants of a given model. During model development, the AI modeler would split their data into multiple parts (i.e., "folds"), repeatedly training and validating their model on different subsets of the dataset. Depending on the modeling technique chosen, the AI modeler might be able to rank the importance, or weight, given to each variable when making a prediction, but it would be quite unlikely that the AI modeler could explain their model in detail. Of course, the AI modeler would need to provide their model as code, as it would not be conveyable in equation form. In return for this compromise, the AI modeler would promise more accurate predictions of lateral spreading displacement. They might prod "but do you *really* understand the Youd et al. (2002) model either, even if it appears in equation form? And regardless, don't you want to make more accurate predictions?" Would the AI modeler be wrong?

**Recurrent Shortcomings (Why Existing AI Liquefaction Models are Ignored)**

Maurer, B. W., & Sanger, M. D. (2023). Why "AI" models for predicting soil liquefaction have been ignored, plus some that shouldn't be. *Earthquake Spectra*, 87552930231173711.

From the preceding background, we next discuss why AI liquefaction models have generally been ignored. This discussion is parsed into five themes.

*Failure to test against SOP liquefaction models*

AI models are widely perceived to be more complex and less transparent than SOP models (in truth AI models are easy to execute, but they are undoubtedly complex in their mathematical underpinnings and opaque in their behaviors). If a practitioner or researcher is to adopt an AI model in lieu of one that is more transparent, trusted, familiar, and widely used, the AI model must be proven. It's therefore perplexing that 67% of the reviewed papers provided no performance comparison against SOP alternatives. Of those that did, some tested against SOP models no longer in common use (e.g., Seed and Idriss, 1971; Stark and Olson, 1995) or appeared to implement SOP models incorrectly (e.g., picking and choosing triggering-model components from different authors). Some papers vaguely allude to performance comparisons without providing any. For example, "The behavior of the proposed neural network model was consistent with prior geotechnical knowledge," and "This rate of successful prediction is about the same or a little higher than those normally achieved by the traditional methods used in practice." Such conclusions are unlikely to be convincing. Other authors establish peculiar measures of success as an apparent defense for foregoing comparisons with SOP models. For example, "Based on accepted geotechnical standards, it was decided that predictions within 50% of the measured horizontal displacements were considered successful." Another group of authors claimed success when their AI model was shown to be 1% more accurate than a different AI model they had previously published. It's worth noting, however, that performance tests against competing models are also rare for SOP models. We are unaware, for example, of any tier-2 SOP liquefaction triggering model that was tested against any other model when originally proposed. The same is true for most tier-2 models that predict manifestations of liquefaction. Such practices are arguably less than ideal but have broadly passed as acceptable in the literature. Of course, SOP models have several advantages in that they: (i) to date have inherently transparent structures; (ii) are generally evolutionary, being relatively incremental improvements on previously trusted methods; and (iii) have been compared and tested against one another in subsequent studies (e.g., Idriss and Boulanger, 2012; Franke et al., 2014; Geyin et al., 2020). None of these mitigating factors presently apply to AI models, making the lack of comparative tests especially problematic. The question thus remains: why should the reader use a new model that hasn't been shown to outperform an existing model that is simpler and more familiar?

*Departure from best practices in model development and performance evaluation*

In addition to lacking tests against SOP models, many AI liquefaction papers depart from three other best practices in model development. First, more than 90% of the reviewed papers quantified model performance using overall accuracy (OA), which is the ratio of accurate predictions to total predictions, and thus varies from 0 to 1. Of the papers that used OA, more than half provided no other measure of model performance. This is generally problematic because OA is highly sensitive to sampling imbalance and liquefaction case-history data is nearly always imbalanced. Consider a hypothetical dataset consisting of 99 "positive" cases (e.g., liquefaction is observed) and 1 "negative" case (e.g., no liquefaction is observed). A hypothetical model that always predicts liquefaction, if applied to this data, would have an OA of 0.99. Although such a model is objectively useless, its OA would suggest it is excellent. It follows that classification models optimized on OA are likely to be biased if trained on imbalanced data. By corollary, biased models can have the illusion of superiority when tested on similarly imbalanced test sets. To be clear, it should not be presumed that every adopter of OA is making a grave mistake. Some authors acknowledge this issue and address it, either by using multiple performance metrics or by balancing the data via resampling. Nonetheless, it's unclear how most models would perform in forward earthquake analyses, given that liquefaction typically occurs at a small fraction of sites where susceptible soils are present, whereas the datasets from which the models are developed are typically imbalanced in the opposite way.

Second, many models exhibit indications of overfitting. Multiple studies, for example, report OAs of 99% or 100% for their AI models. Those who have rigorously studied liquefaction know that it's extremely complex, both in its occurrence and its manifestation. Recent analyses of large global case-history datasets (e.g., Geyin et al., 2020; Rateria and Maurer, 2022) show that tier-2 SOP liquefaction models tend to exhibit efficiencies in the range of 70-85%, albeit there are multiple ways to quantify efficiency and these efficiencies may be higher or lower in individual earthquakes. So, while the SOP models tend to be nearer to perfect models than to random guesses, they are occasionally nearer to random guesses and relatively far from perfection. It's therefore highly suspicious that AI models could be 100% accurate using the same inputs as tier-2 SOP models. Upon investigation, some of the papers reporting high efficiencies do not split their data into training and test sets and do not perform cross-validation, meaning that the model is most assuredly overfit. It's relatively easy to achieve perfection in such cases, since the model is being asked to make predictions only on data it has seen before, and simultaneously, is permitted to be very complex. AI models, like high-order polynomial equations, can segregate two classes of datapoints very efficiently, but this does not mean that either will perform well on unbiased data. Readers should be skeptical of reported efficiencies that greatly exceed those cited above.

Third, several different but related problems stemming from the use of small datasets were observed. Plotted in Figure 4 is a histogram of the number of data points used by each paper to develop an AI model (most often these are data from sites with and without observed liquefaction). Roughly 50% of the reviewed models were developed from less than 250 data points, while 24% used fewer than 150 and 11% used fewer than 75. For comparison, AI models in other fields, such as natural language processing, have been trained on datasets containing 300 billion targets (Hughes, 2023). This provides another possible explanation for the high efficiencies reported by some liquefaction papers. Even if overfitting is properly mitigated, it must be asked whether the test set provides a meaningful challenge for the model, particularly if the dataset is small. Most commonly, 10-20% of a dataset is reserved for unbiased testing, meaning that the number of datapoints used to test a model may be on the order of just 10 to 40. These datapoints may be very similar to datapoints in the training set or otherwise unchallenging to predict. Regardless, performance metrics are only meaningful when multiple models are compared (i.e., SOP models could also be 100% efficient on these test sets).

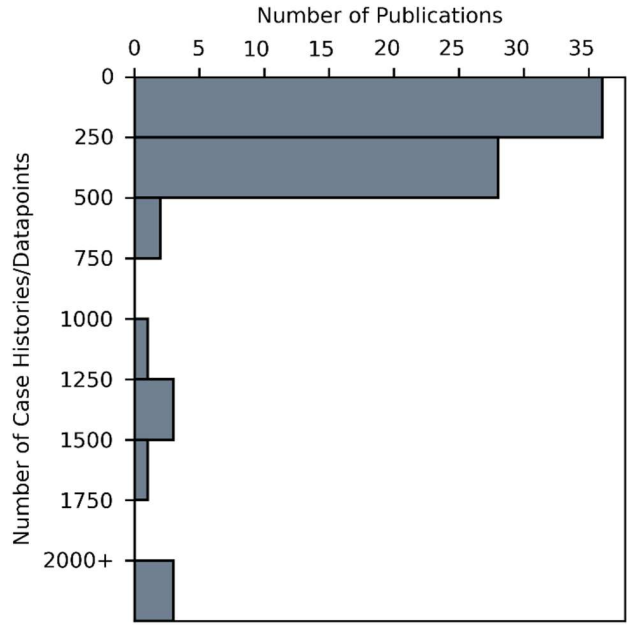

**Figure 4.** Number of liquefaction case histories or datapoints used by the reviewed papers.

The use of small datasets also raises the question of whether measured differences between models are statistically significant, given that any performance metric is likely to have relatively large finite-sample uncertainty. Suppose that 40 liquefaction case histories comprise the test set for comparing multiple models. If these 40 cases were hypothetically selected from a much larger population of case histories, then different samples of 40 cases would surely result in different measurements of performance, and potentially, different conclusions about model superiority. In this way, performance metrics are more uncertain when based on a small sample of the population, all else being equal, and there is less confidence that any one sample (e.g., an observed difference in model performance) accurately reflects the overall population. Yet of the 75 papers reviewed, just one tested for statistical significance. Meanwhile, analyses of tier-2 SOP liquefaction models have shown that measured differences between models are rarely significant in individual earthquakes (Rasanen et al., 2022), or even within global compilations of several hundred case histories (Geyin et al., 2020). Unless proven otherwise, presented differences between AI models and SOP models are likely too small and/or too uncertain to be statistically significant.

Finally, it's noted that blame for using small datasets should not be cast on the respective modelers. Aside from the 2010-2011 Canterbury earthquake sequence, from which a very large quantity of data was recently compiled in the city of Christchurch (e.g., Geyin et al., 2021), only a few hundred "high-quality" case histories have been compiled from all other earthquakes combined. The Boulanger and Idriss (2016) CPT-based case history database, for example, contains 253 cases. The recently expanded Ilgac et al. (2022) SPT database contains 405 cases. And the Kayen et al. (2013) $V_s$ catalog contains 422 cases. The number of measured lateral spreading displacements (e.g., Youd et al., 2002) is on the same order of magnitude. Although the ongoing Next Generation Liquefaction Project (Brandenberg et al., 2020) may increase the size of these datasets, they are unlikely to soon grow substantially, owing to the breadth and expense of data that comprise a single case history. The historic trend, for example, is for case histories to be compiled from less than one earthquake annually. Nonetheless, it must be emphasized that the reviewed papers have generally applied different algorithms to the same small case-history datasets. Many of these papers fail to address or acknowledge the resulting limitations. As seen in Figure 4, a few papers *do* use much larger datasets. Two of these used the newly available Canterbury data, but others did not, which raises another question: how have thousands of data points been used to develop AI liquefaction models when thousands of liquefaction case histories do not exist? This leads to the next thematic shortcoming.

*Use of AI in ways that may not be useful*

In any AI liquefaction paper, it must be asked "what *exactly* is AI being used for, and how might it provide an advantage?" Granting that no two papers are the same, we observe three general uses of AI that warrant critical discussion. Each involves tier-2 triggering models. First, ~7% of the models were not trained on case histories, but rather, were trained to predict the predictions of SOP models. This explains how some papers used very large datasets. With this approach, a set of SPTs or CPTs is typically obtained, and at each depth of measurement an SOP model is used to predict the factor of safety against triggering assuming some seismic loading. Using the SOP model's inputs, an AI model is then trained to predict the SOP model's predictions. It is interesting, perhaps, that the algorithm can learn to reassemble the components of a model to hit a target prediction, but the utility of doing so is unclear. The AI model still requires the same inputs as the SOP model and will never be better than the model it's trained to mimic. This also provides yet another explanation for some of the very high prediction efficiencies in the literature. That is, some models are tested only on their ability to mimic an SOP model and not on their ability to predict liquefaction, yet SOP models are both statistically and conceptually far from perfect (e.g., Geyin et al., 2020; Rasanen et al., 2022; Upadhyaya et al., 2023).

Second, ~13% of the papers propose new limit-state triggering curves, typically in the traditional space of CSR vs. CRR. Any such curve aims to efficiently separate "positive" and "negative" datapoints, or combinations of loading and resistance that are and are not expected to trigger liquefaction, as illustrated conceptually in Figure 5. This use of AI is questionable, however, given that the optimization of a triggering curve can be achieved by simple regression, or even "by eye." As one paper that developed a neural network model concluded: "the limit state function obtained from this study does not seem to be better than the one obtained by simply drawing a boundary curve between the liquefied and non-liquefied data points." Additionally, AI is generally most advantageous when provided with many predictor variables. Fitting an equation to AI predictions is attractive in that it lessens the need to transfer and run code. But to reduce liquefaction to a bivariate curve or trivariate surface (as one paper did) is to operate in the domain of traditional regression, rather than in the wheelhouse of AI.

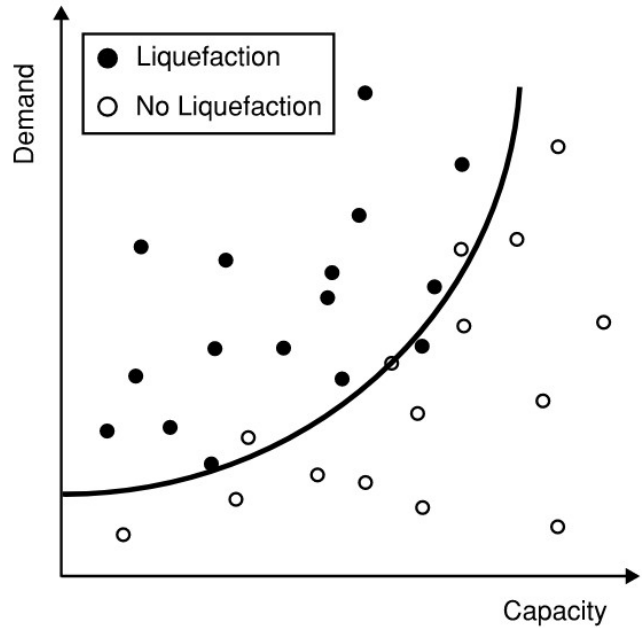

**Figure 5.** Conceptual illustration of limit-state liquefaction triggering curve and data. Is AI needed to separate these two classes of data?

Third, ~55% of the papers develop AI triggering models using the same raw ingredients as tier-2 SOP models (e.g., CPT tip resistance ($q_c$) and sleeve friction ($f_s$), peak ground acceleration (*PGA*), rupture magnitude ($M_w$), depth below ground, fines content (*FC*), total stress ($\sigma$), and effective stress ($\sigma'$)). On one hand, this is a reasonable use of algorithmic learning (i.e., providing AI with numerous predictors thought to be influential, but where the most optimal arrangement for predicting liquefaction is unknown). Simultaneously, the reduction of the simplified stress-based framework to its most raw ingredients is questionable in the context of limited data. Consider, as one example, the stress-reduction coefficient ($r_d$) that is integral to all stress-based triggering models and which predicts CSR at depth as a ratio of CSR at the surface. Models for predicting $r_d$ (e.g., Cetin, 2000; Lasley et al., 2016) are trained independently of liquefaction models using the results of wave-propagation-based site-response analyses of numerous soil profiles and input motions. The $r_d$ model of Green et al. (2020), for example, was trained on the results of 422,400 site-response analyses. Embedded in these models are human knowledge of wave-propagation mechanics and empirical insights from many thousands of sites and motions. To remove $r_d$ is to remove knowledge of mechanics and ask that AI "rediscover" those mechanics by learning from liquefaction case histories. This is not impossible in and of itself, and indeed, our knowledge and modeling of site response are surely imperfect. If provided enough liquefaction observations, AI would learn that CSR varies as a function of depth and other variables. But should we expect AI to learn more about $r_d$ from several hundred liquefaction case histories than we have learned from hundreds of thousands of site-response analyses?

Moreover, AI is being asked not only to understand from liquefaction data the influence of depth, but simultaneously, to understand the influences of groundwater depth, $q_c$, $f_s$, $PGA$, $M_w$, $FC$, $\sigma$, $\sigma'$, etc., and to do so from only a few hundred data points. It is arguably better to isolate and study influential factors like $r_d$ aided with additional data, supplemental analyses, and knowledge of mechanics. Knowledge obtained independent of liquefaction case histories is also embedded elsewhere in the stress-based modeling framework (e.g., the magnitude-scaling, fines-content correction, and overburden correction factors). It's impressive, no doubt, that AI produces apparently reasonable models without these human insights and with limited data, but this does not mean that AI is being used optimally or innovatively. As argued by Geyin et al. (2020), tier-2 triggering models are unlikely to improve greatly with incremental growth in training data or by fine-tuning model parameters, which over the last two decades has produced relatively imperceptible improvements to prediction efficiencies. We would add that this is also likely true for AI, at least insofar as its use to date. AI may indeed play a role in the improvement of these models, but likely only in combination with other disruptive innovations (e.g., new in-situ tests on which models are based, more efficient and sufficient predictor variables, or a fundamental change to the modeling framework, such as that discussed by Geyin et al. (2020) and demonstrated by Upadhyaya et al. (2023)). Outside of these and other new innovations, it may be possible to improve tier-2 triggering models by applying AI to individual model components (e.g., $r_d$), given their semi-empirical nature and the very large quantity of data available to learn from. Our perspective should not be interpreted as opposition to the use of AI by tier-2 developers. Rather, we believe the current status quo, wherein AI is provided a relatively small set of case histories and the conventional predictor variables of SOP models, is unlikely to produce significant improvement.

Additionally, we observed several other decisions and outcomes that the experienced liquefaction researcher would find confusing. One paper developed an AI model considering just two predictor variables: $q_c$ and $PGA$. Another model was shown to perform worse than the SOP model it was being compared to. And a very recent paper advertised the novelty of an "explainable" AI model using 13 predictor variables. The most significant finding from this exercise was that the two most influential variables were $q_c$ and $PGA$. Another similar paper discovered that liquefaction strongly depends on SPT blow count and $FC$. Readers should be suspicious of papers that do not explicitly discuss how and why AI could be advantageous. Is the approach used to develop a model fundamentally different and potentially useful? Does the model outperform existing models? Does the model say anything new about the nature of liquefaction? One or more of the answers arguably ought to be "yes."

### *The "woo-woo" effect*

Papers describing the development of AI liquefaction models tend to be mathematically dense and/or filled with jargon that is unnecessarily involved, and potentially incomprehensible, for non-specialists. It is problematic, of course, when the intended end user cannot understand the paper. One representative abstract developed a "hybrid model using an imperialistic competitive metaheuristic algorithm incorporated with multi-objective generalized feedforward neural network for the purpose of liquefaction potential analysis." This sentence is in no way flawed, but do geotechnical engineers understand its meaning and significance? With each AI technique having multiple hyperparameters, and with various ways to sample datasets, optimize predictors, and ensemble models, there is a seemingly endless array of AI workflows. Each of these variations could be called a novel algorithm and given a unique name, even if AI is generally being applied to liquefaction case-history data in fundamentally the same way. This modeling jargon potentially gives the impression that something newer and/or more innovative than actual is taking place. Of the reviewed papers, we encountered algorithms that collectively were said to be inspired by honeybee swarms, fireflies, ant colonies, the predatory instincts of antlion larvae, and the hunting behaviors of grey wolf packs. Others were inspired by "adaptive resonance theory," utilized

"canonical correlation forests" and "neurofuzzy systems," and had important features such as "t-distributed stochastic neighbor embedment." The adopted algorithms surely have innovative roots and uses in the realm of computer science, yet a simple truth remains: AI liquefaction models will not be used unless geotechnical engineers are convinced of their benefits. To that end, the complexity of language is exacerbated by the fact that AI models are often not well understood even by developers, owing to the innate opacities of AI methods. This too contributes to AI hesitancy. The aura of AI is alluring, but from a pragmatic perspective, model adoption is surely influenced by: (i) a user's familiarity with how a model works and their ability to understand model predictions; and (ii) the perception and acceptance of the model by the broader community (i.e., practicing engineers, regulatory agencies, and others viewed as experts in the field). By contrast, SOP models satisfy each of these criteria, even if acknowledged to be imperfect. AI models are unlikely to become significantly more transparent in the near-term, but the language describing them can and should be better crafted to educate and convince the primary userbase of their utility.

### *Failure to provide the model*

As discussed, and with few exceptions, AI models must be provided in the form of code or embedded in user interfaces, or else no one can use them. It is thus confounding that AI liquefaction models are rarely provided. Of the 75 papers reviewed, just 10 provide a model to readers. Two others state that models will be provided "upon reasonable request," and yet upon reasonable request no models were provided. It is the norm, rather than the exception, for papers to describe the development and performance of an AI liquefaction model without providing the model itself. We concede that this practice may not be unreasonable in all cases. For example, where research is novel and exploratory, it may be reasonable to establish the potential of a new modeling approach without providing a usable model. Similarly, if a study shows that a new modeling approach does *not* work, which is also valuable, then it is reasonable for the authors not to endorse a new model. Still, it is surprising that this is the case in 87% of reviewed papers and hard to imagine in other, more familiar contexts (e.g., those of theoretical laws or traditional regression models). Imagine Youd et al. (2002) describing the development and performance of their model for lateral spreading displacement (i.e., Eq. 1) without ever defining the model itself. Readers would confidently assume that a mistake had occurred during the publication process, and yet this is normal in the context of AI liquefaction models. We surmise that editors and reviewers may be insufficiently familiar with AI tools, or understandably disoriented by the woo-woo effect, to recognize that papers are describing models that cannot be used by anyone other than the authors. Whatever the case may be, it is unsurprising that models which are not provided to readers will not be used.

Shown in Figure 6 is a Sankey diagram depicting the prevalence of different features for the 75 reviewed papers. Namely: (i) the type of AI tool used; (ii) the type of liquefaction model developed; (iii) whether any comparison to an SOP model was made; and (iv) whether the code was provided (i.e., either raw or in a user interface). Most papers used a neural network algorithm; developed a tier-2 liquefaction model; did not compare to any SOP model; and did not provide the model to readers. Additionally, most papers are deploying AI in ways that differ only slightly, on the same small case-history datasets, for purposes that we believe are unlikely to produce significant gains over SOP models. If a new, more opaque model isn't shown to outperform a simpler, more trusted model and isn't even provided for use, might readers be justified in ignoring the work?

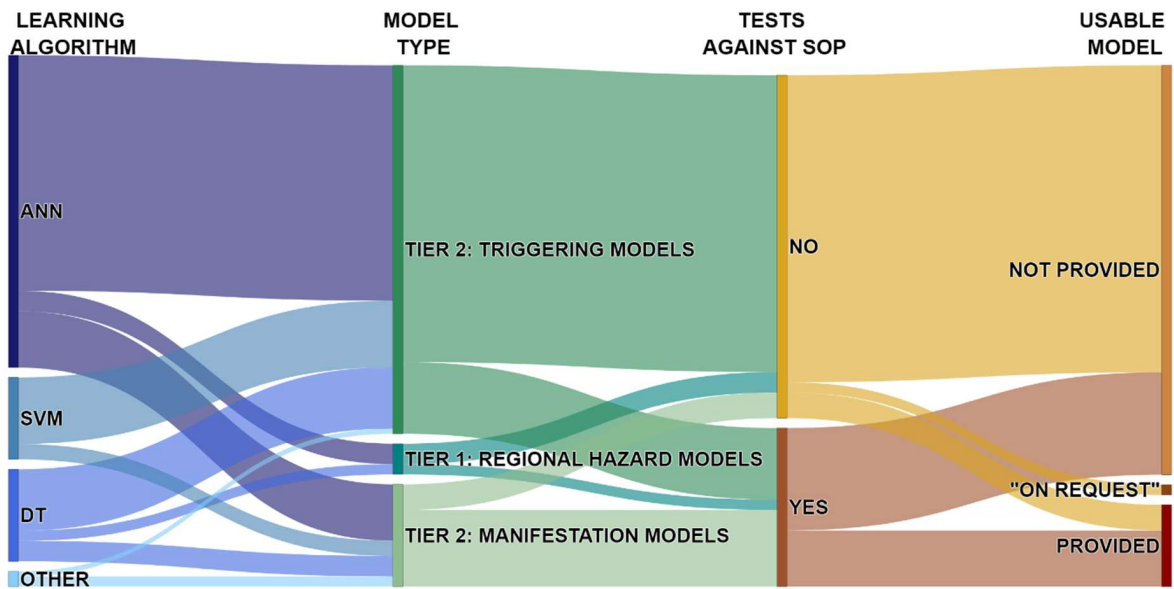

**Figure 6.** Sankey diagram illustrating the prevalence of model features; ANN = artificial neural network; SVM = support vector machine; DT = decision tree.

**Applications and Models that Shouldn't be Ignored**

While ~75% of reviewed papers developed tier-2 triggering models, we observed a general trend amongst papers that developed either tier-1 "geospatial" models or tier-2 manifestation models. Namely, these papers were more likely to be free from the thematic shortcomings discussed above. More broadly, we believe AI is best used where predictors and a target response are correlated but not mechanistically linked (e.g., when predicting below-ground conditions from above-ground parameters) or where predictors and a response should seemingly be explainable by mechanics, but those mechanics have not yet been well modeled (e.g., when predicting the relationship between liquefaction at depth and its manifestation at the ground surface). In the following, each of these research areas is briefly expanded on and representative papers are highlighted.

*Tier-1 Geospatial Models*

In a perfect world, liquefaction would be rapidly predicted at high spatial resolution across the regional extents of large earthquakes. This seemingly remains a fantasy, however, given that tier-2 and 3 liquefaction models require in-situ test data, which cannot feasibly be collected continuously across vast areas. Interest has thus grown in geospatial models, which are well suited for regional-scale applications, and which have recently been adopted by agencies in the European Union and United States (e.g., Lai et al., 2019; Allstadt et al., 2021). The U.S. Geological Survey, for example, utilizes the seminal model of Zhu et al. (2017) in scenario and post-event data products. This model is a regression equation with five variables. One represents demand (peak ground velocity) and four represent capacity (distance to water, annual precipitation, groundwater depth, shear wave velocity time-averaged over the upper 30 m).

It is inherently challenging to predict subsurface conditions without subsurface tests, especially across diverse geomorphic, topographic, and climatic environs. This is an ideal problem for AI, given that subsurface properties lack mechanistic links to geospatial variables, but correlate in complex, interrelated ways. In this regard, AI might: (i) provide learning insights that are unlikely with traditional regression, given the number of viable predictor variables; (ii) allow for more predictive information to be used, with greater potential for that information to be exploited; and (iii) produce models via ensembling that are more generalizable across environments. Given this motivation, Todorovic and Silva (2022) developed a global model in the general style of Zhu et al. (2017) using six predictors and a decision-tree ensemble algorithm. Todorovic and Silva (2022) used cross-validation during model training and offered evidence of improvement over Zhu et al. (2017) via unbiased testing. Todorovic and Silva (2022) also stated that software for running the model will be incorporated in the OpenQuake Engine (e.g., Pagani et al., 2014), albeit we were unable to locate that software at the time of our writing.

In parallel, Geyin et al. (2022) developed an AI geospatial model using 12 variables and a fundamentally different approach. Whereas all other geospatial models have trained on outcomes (i.e., observations of liquefaction), Geyin et al. (2022) proposed training on the mechanistic causes of those outcomes (i.e., subsurface engineering properties). In effect, this model predicts subsurface CPT data within the context of tier-2 liquefaction triggering and manifestation models. Geyin et al. (2022) argued this has two main advantages. First, it transfers the AI prediction target from liquefaction observations (for which the training set is relatively small and slow to grow) to subsurface measurements (for which the potential training set is vast, since the sites of in-situ tests need not have experienced an earthquake). Second, it retains the knowledge of liquefaction mechanics that has developed over the last 50+ years and which is embedded in tier-2 models, thereby providing a mechanistic underpinning to ensure sensible response and scaling. This approach thus parses the problem into that which is empirical and arguably best predicted by AI (the relationship between geospatial variables and subsurface traits) and that which is arguably best predicted by mechanics (liquefaction response, conditioned on those traits). The resulting model was trained on just 1,700 CPTs but was shown – to a statistically significant degree – to perform at least as well as Zhu et al. (2017). It was also coded in new Windows software, *RapidLiq*, making its implementation trivial. The Geyin et al. (2022) model provides one example of how AI and mechanics might exist harmoniously, with AI working to understand empirical patterns within the confines of an overarching mechanical framework.

Maurer, B. W., & Sanger, M. D. (2023). Why "AI" models for predicting soil liquefaction have been ignored, plus some that shouldn't be. *Earthquake Spectra*, 87552930231173711.

In addition to these efforts, authors of the seminal Zhu et al. (2017) model have also preliminarily explored the use of AI (e.g., Asadi et al., 2022). Other tier-1 geospatial models will surely be forthcoming.

*Tier-2 Manifestation Models*

In contrast to tier-2 models for predicting liquefaction triggering, those for predicting manifestations (e.g., lateral spreading or ejecta) have generally used methods based more on intuition and empiricism, owing to the complexity of these multidimensional hydromechanical phenomena. Although efforts to understand manifestation mechanics have increased in the wake of the Canterbury sequence and should continue (e.g., Cubrinovski et al., 2019; Bassal and Boulanger, 2021; Hutabarat and Bray, 2022), these efforts have either not yet produced new tier-2 models, or the resulting models have not yet been shown to outperform existing methods. These efforts have also highlighted the potential advantages of tier-3 models for producing insights into manifestation patterns and severities, and thus, may instead benefit the profession by increasing the use of more advanced methods. Still, the insights gained from these efforts should be reducible and transferable to simpler models. Alas, the knowledge gap between triggering and manifestation is diminishing but persistent.

This gap has long invited empiricism (e.g., see Eq. 1), for which AI is well suited. Oommen and Baise (2010), for example, developed an AI model for predicting lateral spreading displacement using the same predictors as Youd et al. (2002). Their model, which used cross-validation during training, was shown to outperform the Youd et al. (2002) SOP model and was provided as code. Liu and Tesfamariam (2012) developed a similar suite of models with many of the same strengths as Oommen and Baise (2010), albeit they did not test against an SOP model. More recently, Durante and Rathje (2021) developed AI models for predicting lateral spreading occurrence and displacement using 7,300 datapoints from the 2011 Christchurch earthquake and a novel combination of CPT and geospatial predictors. Although further work is needed to develop models generalizable to other events and settings, Durante and Rathje (2021) employed cross-validation, used an unbiased test set, and provided their model as code. Lastly, Rateria and Maurer (2022) used 14,400 data points to develop an AI updating of the Ishihara (1985) "$H_1$-$H_2$" model for predicting liquefaction surface expression. This model adhered to best practices, outperformed competing models in unbiased testing, and was provided as code. Simultaneously, this work demonstrated the insufficiency of the $H_1$-$H_2$ model variables and the role that mechanical knowledge must play in generating better predictors. To that end, the tier-2 manifestation model of Hutabarat and Bray (2022) uses novel and insightful predictors inspired by tier-3 modeling results, but ultimately, is resigned to conventional empiricism when developing and calibrating a model. This is another example of how AI might improve the empirical component(s) of a model while also respecting and benefiting from human knowledge of mechanics. No matter how complex future prediction models become, it is likely they will continue to be semi-empirical, and therefore, potentially improved by AI.

While it remains to be seen whether any of the highlighted models will be widely used, they have generally adhered to best practices in model training and have generally been shown to outperform SOP counterparts. They arguably also demonstrate reasonable and useful applications of AI. Most importantly, these models have been provided to readers so that others may test them using new data and report on their performances. The same cannot be said for most AI liquefaction models.

**Conclusions**

We have argued, via a review of 75 publications, that there are valid reasons why existing AI-based models for predicting soil liquefaction have not been adopted for use. This should not be interpreted to mean that prior efforts have been without merit or utility. Rather, this is the result of our investigation of a simple fact: almost no one is using these models other than their respective developers. Simultaneously, we believe that AI can, and very likely will, play an important role in the field, just as it increasingly does in most fields of empirical inquiry. Inevitably, that future will have been built on strides collectively made by the current AI liquefaction literature. We have discussed the prevailing and often severe problems with much of the current literature – recognition of which is inherently easier in retrospect – so that these problems may be understood, identified, and remedied. Our intention is to shape the direction and perceptions of this growing body of work for the better. Towards that end, we also highlighted select papers generally free from the recurrent shortcomings discussed herein. These papers demonstrate areas of research where AI is more likely to provide value in the near term: permitting new modeling approaches and potentially producing more accurate predictions of liquefaction phenomena.

**Data Availability Statement**

All data, models, and code generated or used during the study appear in the submitted article.

**Acknowledgements**

The presented study is based upon work supported by the US Geological Survey (USGS) under Grant No. G23AP00017. However, any opinions, findings, and conclusions or recommendations expressed in this paper are those of the authors and do not necessarily reflect the views of USGS.

Maurer, B. W., & Sanger, M. D. (2023). Why "AI" models for predicting soil liquefaction have been ignored, plus some that shouldn't be. *Earthquake Spectra*, 87552930231173711.

Bassal, P.C., and Boulanger, R.W. 2021. "System response of an interlayered deposit with spatially preferential liquefaction manifestations." *Journal of Geotechnical and Geoenvironmental Engineering*, 147(12), 05021013. https://doi.org/10.1061/(ASCE)GT.1943-5606.0002684.

Baziar, M.H. and Ghorbani, A. 2005. "Evaluation of lateral spreading using artificial neural networks." *Soil Dynamics and Earthquake Engineering*, 25(1), 1-9. https://doi.org/10.1016/j.soildyn.2004.09.001.

Baziar, M. H. and Nilipour, N. 2003. "Evaluation of liquefaction potential using neural-networks and CPT results." *Soil Dynamics and Earthquake Engineering*, 23(7), 631-636. https://doi.org/10.1016/S0267-7261(03)00068-X.

Boulanger, R.W. and Idriss, I.M. 2014. "CPT and SPT based liquefaction triggering procedures." *Report No. UCD/CGM-14, Center for Geotechnical Modeling, Department of Civil and Environmental Engineering, University of California, Davis, CA,* 1-134. Available at: https://ucdavis.app.box.com/s/vqgqjyvyby9w4xpegk7av2yu2ytact64.

Boulanger, R.W. and Idriss, I.M. 2016. "CPT-based liquefaction triggering procedure." *Journal of Geotechnical and Geoenvironmental Engineering*, 142(2), 04015065-04015065. https://doi.org/10.1061/(ASCE)GT.1943-5606.0001388.

Brandenberg, S.J., Zimmaro, P., Stewart, J.P., Youp Kwak, D., Franke, K.W., Moss, R.E., Cetin, K.O., Can, G., Ilgac, M., Stamatakos, J., Weaver, T., and Kramer, S.L. 2020. "Next-generation liquefaction database." *Earthquake Spectra*, 36(2), 939-959. https://doi.org/10.1177/8755293020902477.

Cetin, K.O. 2000. "Reliability-based assessment of seismic soil liquefaction initiation hazard." Doctor of Philosophy Dissertation, Univ. of California, Berkeley, CA.

Cetin, K.O., Seed, R.B., Der Kiureghian, A., Tokimatsu, K., Harder Jr, L.F., Kayen, R.E., and Moss, R. E. 2004. "Standard penetration test-based probabilistic and deterministic assessment of seismic soil liquefaction potential." *Journal of Geotechnical and Geoenvironmental Engineering*, 130(12), 1314-1340. https://doi.org/10.1061/(ASCE)1090-0241(2004)130:12(1314).

Cetin, K. O., Seed, R. B., Kayen, R. E., Moss, R. E. S., Bilge, H. T., Ilgac, M., and Chowdhury, K. 2018. "SPT-based probabilistic and deterministic assessment of seismic soil liquefaction triggering hazard." *Soil Dynamics & Earthquake Engineering,* 115, 698-709. https://doi.org/10.1016/j.soildyn.2018.09.012.

Chen, C.H., Juang, C.H., and Schuster, M.J. 2008. "Empirical model for liquefaction resistance of soils based on artificial neural network learning of case histories." In *Proc., GeoCongress*, 854-861. New Orleans: Characterization, Monitoring, and Modeling of GeoSystems. https://doi.org/10.1061/40972(311)107.

Chern, S.G. and Lee, C.Y. 2009. "CPT-based simplified liquefaction assessment by using fuzzy-neural network." *Journal of Marine Science and Technology*, 17(4), 326-331. https://doi.org/10.51400/2709-6998.1990.

Chern, S.G., Lee, C.Y., and Wang, C.C. 2008. "CPT-based liquefaction assessment by using fuzzy-neural network." *Journal of Marine Science and Technology*, 16(2), 139-148. https://doi.org/10.51400/2709-6998.2024.

Chiru-Danzer, M., Juang, C.H., Christopher, R.A., and Suber, J. 2001. "Estimation of liquefaction-induced horizontal displacements using artificial neural networks." *Canadian Geotechnical Journal*, 38(1), 200-207. https://doi.org/10.1139/t00-087.

Cubrinovski, M., Rhodes, A., Ntritsos, N., and Van Ballegooy, S. 2019. "System response of liquefiable deposits." *Soil Dynamics and Earthquake Engineering*, 124, 212-229. https://doi.org/10.1016/j.soildyn.2018.05.013.
31

Maurer, B. W., & Sanger, M. D. (2023). Why "AI" models for predicting soil liquefaction have been ignored, plus some that shouldn't be. *Earthquake Spectra*, 87552930231173711.

Kumar, V. and Rawat, A. 2017. "Soil liquefaction and its evaluation based on SPT by soft-computing techniques." MATTER: International Journal of Science and Technology, 3(2), 316-327. https://doi.org/10.20319/mijst.2017.32.316327.

Kurnaz, T. F. and Kaya, Y. 2019. "A novel ensemble model based on GMDH-type neural network for the prediction of CPT-based soil liquefaction." *Environmental Earth Sciences*, 78(11), 1-14. https://doi.org/10.1007/s12665-019-8344-7.

Kurup, P. U. and Garg, A. 2005. "Evaluation of liquefaction potential using neural networks based on adaptive resonance theory." *Transportation Research Record*, 1936(1), 192-200. https://doi.org/10.1177/0361198105193600122.

Lai, C. G., Conca, D., Famà, A., Özcebe, A. G., Zuccolo, E., Bozzoni, F., Meisina, C., Bonì, R., Poggi, V., Cosentini, R. M. 2019. "Mapping the liquefaction hazard at different geographical scales." In *Earthquake Geotechnical Engineering for Protection and Development of Environment and Constructions*, (Eds. F. Silvestri and N. Moraci), pp. 691-704. CRC Press. https://doi.org/10.1201/9780429031274.

Lasley, S. J., Green, R. A., and Rodriguez-Marek, A. 2016. "New stress reduction coefficient relationship for liquefaction triggering analyses." *Journal of Geotechnical and Geoenvironmental Engineering*, 142(11), 06016013. https://doi.org/10.1061/(ASCE)GT.1943-5606.0001530.

Liu, B. Y., Ye, L. Y., Xiao, M. L., and Miao, S. 2006. "Artificial neural network methodology for soil liquefaction evaluation using CPT values." In *International Conference on Intelligent Computing,* 329-336. Springer, Berlin, Heidelberg. https://doi.org/10.1007/11816157_36.

Liu, Z. and Tesfamariam, S. 2012. "Prediction of lateral spread displacement: data-driven approaches." *Bulletin of Earthquake Engineering*, 10(5), 1431-1454. https://doi.org/10.1007/s10518-012-9366-7.

Livingston, G., Piantedosi, M., Kurup, P., and Sitharam, T. G. 2008. "Using decision-tree learning to assess liquefaction potential from CPT and Vs." In *Proc. of the Geotechnical Earthquake Engineering and Soil Dynamics IV*, 1–10. Sacramento. https://doi.org/10.1061/40975(318)76.

Maurer, B. W., Green, R. A., and Taylor, O. D. S. 2015. "Moving towards an improved index for assessing liquefaction hazard: Lessons from historical data." *Soils and Foundations*, 55(4), 778-787. https://doi.org/10.1016/j.sandf.2015.06.010.

Moss, R., Seed, R., Kayen, R., Stewart, J., Kiureghian, A., and Cetin, K. E. M. A. L. 2006. "CPT-based probabilistic and deterministic assessment of in situ seismic soil liquefaction potential." *Journal of Geotechnical and Geoenvironmental Engineering*, 132(8). https://doi.org/10.1061/(ASCE)1090-0241(2006)132:8(1032).

Muduli, P. K. and Das, S. K. 2014. "Evaluation of liquefaction potential of soil based on standard penetration test using multi-gene genetic programming model." *Acta Geophysica*, 62(3), 529-543. https://doi.org/10.2478/s11600-013-0181-6.

Nejad, A. S., Güler, E., and Özturan, M. 2018. "Evaluation of liquefaction potential using random forest method and shear wave velocity results." In *Proc. Of the International Conference on Applied Mathematics & Computational Science (ICAMCS. NET)*, 23-233. Budapest: IEEE. https://doi.org/10.1109/ICAMCS.NET46018.2018.00012.

Njock, P. G. A., Shen, S. L., Zhou, A., and Lyu, H. M. 2020. "Evaluation of soil liquefaction using AI technology incorporating a coupled ENN/t-SNE model." *Soil Dynamics and Earthquake Engineering*, 130, 105988. https://doi.org/10.1016/j.soildyn.2019.105988.
35